\documentstyle[aps,psfig,multicol]{revtex}
\begin{document}
\topmargin -15mm

\title{Magnetism, Spin-Orbit Coupling, and Superconducting Pairing in UGe$_2$}

\author{A. B. Shick and  W. E. Pickett}

\address{Department of Physics, University of California, Davis, CA 95616}

\maketitle

\begin{abstract}
A consistent picture on the mean-field level of the magnetic properties and
electronic structure of the superconducting itinerant ferromagnet UGe$_2$ is
shown to require inclusion of correlation effects beyond the local density
approximation (LDA).  The ``LDA+U" approach
reproduces both the magnitude of the observed moment, composed of
strongly opposing spin and orbital parts, and the magnetocrystalline
anisotropy.  The largest Fermi surface sheet is comprised primarily of 
spin majority states with orbital projection $m_{\ell}$=0, 
suggesting a much simpler picture of 
the pairing than is possible for general strong spin-orbit 
coupled materials.  This occurrence, and the quasi-two-dimensional geometry
of the Fermi surface, support the likelihood of
magnetically mediated p-wave triplet pairing.
\end{abstract}

\begin{multicols}{2}
The coexistence of superconductivity with magnetism has recently re-emerged as a
central topic in condensed matter physics, due to competition between 
ordering of magnetic ions and superconducting pairing in 
borocarbides \cite{borocarb}, to unprecedented
coexistence of magnetic order with substantial ferromagnetic component
with high-temperature 
conductivity \cite{Tallon,Pickett,Lynn}
and to observation of magnetic $\rightarrow$ superconducting 
transitions near quantum critical
points in f electron systems such CeRh$_2$Si$_2$\cite{Movsh}, CeIn$_3$
\cite{Walker}, CePd$_2$(Si,Ge)$_2$ \cite{Lister}, and CeRhIn$_5$
\cite{Hegger}.
Superconductivity
below 1K  in the limited pressure range (1 to 1.6 GPa) 
was recently observed coexisting with strong ferromagnetism
in a high purity single crystal UGe$_2$ \cite{Lonzarich,Saxena}
adding yet another dimension to this unanticipated phenomenon of
phase coexistence.
This superconducting phase is found {\it within} the ferromagnetic phase and
disappears in the paramagnetic region, strongly suggesting the pairing
mechanism is magnetic in origin.  

There are no other known examples of superconductivity coexisting with 
{\it strong} ferromagnetism (average moment comparable to Ni, say).  This 
phenomenon necessarily requires triplet spin pairing (or more precisely,
an odd symmetry real space pair) since the exchange split bands (and the
very different Fermi surfaces) preclude singlet pairing.  The normal state
symmetry itself is low in spite of the relatively simple crystal structure
of UGe$_2$.  Magnetism breaks time reversal invariance, leaving the normal
state symmetry consisting of the product of the magnetic space group 
and gauge invariance [${\cal G} \times U(1)$], the latter of which is
always broken by the onset of superconductivity.  The possible order
parameter symmetries are very limited.

Recent experiments on single crystals\cite{Oikawa,Boulet} indicate UGe$_2$
(previously thought to crystallize in the $Cmcm$ ZrSi$_2$ type structure)
actually has the base-centered
orthorhombic ZrGa$_2$ crystal structure ($Cmmm$). The structure, shown in
Fig. 1, can be viewed as consisting of antiphase zigzag chains of U atoms
running along the $\hat a$ direction and lying within the $\hat a - 
\hat b$ plane; however, interchain and intrachain separations are
comparable.  Each U is tenfold coordinated by Ge.  Importantly, the 
structure possesses inversion symmetry.  Single crystal
magnetization measurements \cite{Menkovsky} and neutron powder diffraction 
measurements \cite{Boulet} both yield a collinear magnetic structure with
the ferromagnetically ordered magnetic moment of 1.42 $\mu_B$.  (We quote
moments per formula unit, {\it i.e.} per U atom.)
The Curie temperature $T_c = 52$ K at ambient pressure decreases under
pressure, vanishing at 1.6 GPa.  Around 1 GPa, 
Saxena {\it et al.}\cite{Saxena} have
found that UGe$_2$ becomes superconducting while remaining strongly
ferromagnetic (${\bar M} \; \approx \; 1~\mu_B$/U) providing a novel
example of coexistence of the 
superconductivity with strong ferromagnetism.

Magnetic measurements  \cite{Menkovsky,Onuki} yield very strong magneto-crystalline anisotropy in UGe$_2$
with easy magnetization axis along $\hat a$ 
(the shortest crystallographic axis, Fig. 1). It was 
found\cite{Menkovsky}  that 
even in high magnetic field (up to 35 T) it is impossible to saturate 
magnetization along the hard $\hat b$ and $\hat c$  axes.
Such highly anisotropic behavior is typical of magnetic $5f$ electron materials. 
Susceptibility measurements\cite{Saxena} above T$_C$ yield an 
effective paramagnetic moment $m_{eff} \approx 2.7 \; \mu_B$/U which 
differs from atomic value of $3.62 \; \mu_B$ for atomic $f^3$ configuration,
indicating substantial 5f hybridization with conduction states. 
Moreover, the carriers with
the high cyclotron masses of 15-25 $m_0$ were observed in dHvA experiments \cite{Onuki2} suggesting itinerant but
strongly correlated 5f-electron states. UGe$_2$ is however not a 
real heavy fermion system since
its electronic specific heat coefficient ${C(T)}/{T}$ = 
$\gamma$ $\approx$ 35 mJ/K$^2$mol is about ten times smaller
than in conventional heavy-fermion U-compounds \cite{Onuki}.

The electronic and magnetic structure of UGe$_2$ is a rather open issue. Yamagami {\it et al.} \cite{Yamagami}
reported the results of relativistic augmented-plane-wave calculations, but without interplay between magnetism and spin-orbit coupling (SOC)
and using the old crystal structure.
The scope of this paper is to provide a microscopic picture of electronic structure
and magnetism  of ferromagnetic UGe$_2$ upon which models of the 
superconductivity can be based.  This picture leads to specific indications of 
the pairing character in UGe$_2$.

We first applied conventional band theoretical methods in the local
density approximation using a relativistic full-potential 
linearized augmented plane wave method
\cite{flapwso}.  The results did not reproduce
the observed ground state moment\cite{Menkovsky,Boulet}. 
The calculated spin magnetic moment ($M_s$ = 1.57 $\mu_B$) is canceled by
the orbital moment arising from spin-orbit coupling ($M_L$ = -1.84 $\mu_B$ \cite{OrbMom}) 
yielding the net moment of $|M|$ = 0.27 $\mu_B$,
five times smaller than the experimental value of 1.4 $\mu_B$.  The cause
is due to the oversimplified treatment of correlation effects, as is often seen in the
application of LDA to f electron materials.
   
We therefore account for the on-site atomic-like correlation effects beyond LDA
by using the LDA+U approach\cite{Sasha} in a rotationally invariant, full potential
implementation.
Minimizing the LDA+U total energy functional
with SOC treated self-consistently\cite{flapwso} generates not only the
ground state energy and spin densities, but also effective one-electron 
states and energies (the band structure) that provides the orbital 
contribution to the moment and Fermi surfaces. The basic difference of
LDA+U calculations from the LDA is its explicit dependence on the
on-site spin- and orbitally resolved occupation matrices \cite{CDF}.
Since the LDA+U method is rarely applied to metals, the appropriate values of
the on-site f electron repulsion $U$ and exchange $J$ constants are not known.
Our values of $U$ = 0.7 eV, $J=0.44$ eV were chosen to reproduce the 
ground state magnetic moment
$M_J \; = \; M_s \; + \; M_l$.
Mainly $U$ was varied, but it seemed to be of some importance to reduce $J$ slightly
from its atomic value of 0.55 eV \cite{Marel} to keep $U$ from becoming unreasonably small.   
The resulting total uranium f-electron occupation (2.8 $e$) is close to but clearly less
than the $f^3$ configuration.     
The value of Coulomb $U$ differs from the ``atomic-like" value
of $U$ = 2 eV which was derived for U $f^3$ compounds in Ref. \cite{Marel}.
The reduction of the Coulomb $U$
to 0.7 eV in order to obtain the correct value for the 
magnetic moment reflects the partly
itinerant character of the U 5f states and accounts for 
the metallic screening of the
U 5f states in UGe$_2$. 

To probe the magnetic ground state we calculated the energy
of one antiferromagnetic configuration in addition to ferromagnetic,
where the two U atoms in the unit cell (in the ``chain'' in Fig. 1)
had antialigned moments.  Nearest neighbors between chains there were
also antiparallel.
The ferromagnetic alignment is 32 meV/f.u. lower in energy, consistent
with the observed ferromagnetism.

To study the magnetic anisotropy we performed
the total energy calculations for the magnetic moment along each of the 
$\hat a$, $\hat b$, and $\hat c$ axes.
The orthorhombic crystal symmetry was reduced in order to preserve the
projection of the orbital moment along the spin direction.
To obtain convergence for the total energy differences we used 
275 k-points in 1/4th of
the BZ for the magnetization along $\hat a$ and $\hat b$, and 
250 k-points for $\hat c$ orientation; these meshes are equivalent to
1000 k-points in the BZ.  The total energy differences 
yield the magnetic anisotropy energy and are shown in Table I. 
The calculations
produce easy magnetization axis $\hat a$ in accord with the experiment.
There is a pronounced anisotropy for the both spin $M_s$ and orbital 
$M_L$ magnetization
reflecting the interplay between the strong spin-orbit coupling 
(the SOC constant $\xi = 0.22$ eV) and the large exchange splitting.
The variation of the spin and orbital moments with orientation 
is 7-10\% (Table I), and
manifests itself in strong, qualitative differences in the band
structure and Fermi surfaces.  Thus it is {\it crucial} to use the easy
axis orientation, and henceforward we discuss only this case.

This strong magnetic anisotropy (2-3 orders of magnitude larger
than in 3d ferromagnets)  reflects the 
strong spin-orbit coupling for 5f-states 
and  is consistent with the reported experimental data for the MAE in the uranium compounds \cite{Divis}.
It explains qualitatively the difference in the measured magnetization
values for single crystals (1.42 $\mu_B$) and polycrystals (1.07 $\mu_B$ \cite{Boulet}, 0.5 $\mu_B$ \cite{Onuki}):
since the magnetic moment is kept by magnetic anisotropy along
the easy magnetization direction ($\hat a$) in each grain of the 
polycrystal, the average measured moment should be reduced but 
should exceed one third of its single crystal value.

The complex band structure contains 7 orbitals $\times$ 2 U atoms $\times$
2 spins = 28 primarily U 5f bands within a 3 eV region spanning 
the Fermi energy E$_F$.
They are of mixed spin character due to SOC and contain only minor Ge
4p character, which is mostly below and above the U 5f bands.
Bands crossing the Fermi level have predominantly U 5f 
character and mostly spin majority character. 
These bands have low dispersion along $k_y$ ($\hat b$ direction) indicating
an $\hat a - \hat c$ plane quasi-two-dimensional (quasi-2D) character 
of the UGe$_2$ electronic structure near 
E$_F$.  The exchange splitting, to
the extent that an average can be defined when there is strong SOC, is
about 20 times as large (1-1.5 eV) as suggested by 
Yamagami {\it et al.}\cite{Yamagami} based on unpolarized calculations.

Since the LDA+U potential is formulated in terms of $m_s$ and $m_l$ 
rather than the total moment $m_j$,
it is natural to perform a spin- and orbital- resolution
$\{ m_s$;$m_l \}$.  The combination of strong spin 
polarization and large SOC result in a remarkably clean $\{ m_s$;$m_l \}$ 
separation as is evident in Fig. 2.
The peak in the vicinity of E$_F$ is formed by 
$\{ \uparrow ; 0 \}$ and $\{ \downarrow ; 1 \}$ states coupled by spin-orbit;
$N(E_F)$/f.u. = 1.7 and 0.7 states/eV respectively.
The total $N(E_F)$/f.u. of
5.5  states/eV is mainly (75-80\%) due to U 5f contributions.
The measured electronic specific heat coefficient 
$\gamma$ = 35 mJ/K$^2$ mol \cite{Onuki} corresponds to
a dressed value $N^*(E_F)$ = 15 states/eV, indicating a 
dynamic enhancement $N^*(E_F)/N(E_F)$= 2.7 arising from magnetic fluctuations
with possible contributions from phonons and charge fluctuations.

Photoelectron (XPS-BIS) studies  of UGe$_2$\cite{Suzuki} reported a 
pronounced broad peak in XPS spectrum $\approx$ 1 eV 
below the Fermi level.  Our calculated 
U fDOS (Fig. 2) has a double peak 
due to $\{ \uparrow; m_l = -3,-2 \}$ located  0.3-0.4 eV 
below the Fermi level.  This difference may be due to neglect of dynamical
effects in our calculations but may also reflect surface state 
contributions in the data.
In the measured BIS spectrum, which is less surface sensitive, 
there is a kink around 0.5 eV and 
a main peak at 1.3 eV above E$_F$. The LDA+U 
calculations reproduce both features: the structure
at 0.5 eV arises from $\{ \uparrow ; -1 \}$ 
character
and the much bigger peak at around 1.5 eV is due to $\{ \uparrow ; +3 \}$ 
and $\{ \downarrow ; 0,1 \}$ states.

The calculated Fermi surface is complex and consists of three sheets. 
The largest and most interesting
sheet is shown in Fig. 3.
The quasi-2D  surface has weak dispersion along $k_y$, {\it i.e.}
hopping between chains is small for this band. 
In spite of (but actually due to) the strong SOC, 
this surface has simple character: predominantly 
$\{ \uparrow ; 0 \}$ with
some mixture of $\{ \downarrow ; 1 \}$.
If, in a first approximation, this small opposite spin
character can be neglected, one can consider pure spin
$\{ \uparrow ; 0 \}$ states and hence simpler models for pairing.
In this limit the problem reduces to the one of so-called ``single spin
pairing'',\cite{Rudd} which however has not been studied specifically 
for the low orthorhombic symmetry of UGe$_2$.

It was suggested some time ago\cite{Fay} that superconducting
p-wave (triplet) pairing for equal spin states can appear
due to the longitudinal magnetic fluctuations.
The roughly nesting portions of quasi-2D surfaces should promote
strong magnetic interactions between spin-majority carriers 
located periodically in the space along magnetization direction $\hat a$.
The rapid collapse of the
moment between 1 GPa and 1.6 GPa\cite{Saxena} is also consistent with
longitudinal magnetic fluctuations.
The distinctive electronic structure that we find
suggests that a non-relativistic model can be used to 
describe the magnetically
mediated superconductivity in UGe$_2$. 

It must be recognized that the results presented here apply only to the zero
pressure and temperature.  At the peak of the superconducting phase 
($\sim$ 1GPa) the moment is reduced by one-third and the lattice constants
and internal coordinates
will be changed.  Consequently the 
exchange splitting will decrease and band widths will be increased, 
altering the Fermi surface from the one we have presented.   The general
features of the ($m_s;m_{\ell}$) fDOS separation in Fig. 2 are expected
to be robust.

To summarize, accounting for some correlated behavior (beyond what is included
in LDA) using the LDA+U 
approach in a very general implementation including SOC, we have 
provided a microscopic picture of the electronic and magnetic character
of UGe$_2$ that is essential for beginning to understand the coexistence
of superconductivity with strong ferromagnetism.  Our results are 
in accord with several important experimental observations (magnitude and
direction of the moment, BIS spectra), and suggest a relatively simple
picture of the U 5f bands arises in spite of the 28 f bands that span
$E_F$.
The calculated Fermi surface has a quasi-2D sheet of predominantly majority
spin character, supporting conjectures of magnetically mediated 
triplet superconductivity in UGe$_2$.  The coexistence of superconductivity
with magnetic order is quite distinct from that in RuSr$_2$GdCu$_2$O$_8$,
with its separate magnetic superconducting layers, because in UGe$_2$ 
the magnetism and superconductivity clearly seem to be due to the
same U 5f electrons.

We are grateful to S.S. Saxena for providing the manuscript \cite{Saxena} prior to publication
and for useful discussion. This work was supported by ONR Grant N0001-97-1-0956 and NSF
Grant No. DMR9802076.


\begin{table}
\caption
{The U 5f state occupation ($n_f$), spin ($M_S$), orbital ($M_L$) and 
absolute value of net ($|M_J|$) magnetic moments ($\mu_B$)
for three directions of magnetization.  The magnetocrystalline anisotropy
energy (MAE in mRy/f.u.) is the total energy difference 
$MAE \; \equiv E_{b(c)} \; - \; E_a$ for the magnetization directed
as indicated.} 
\begin{tabular}{cccccccc}
       &$n_f$  & $M_S$ & $M_L$ & $|M_J|$ & $MAE$ \\
$\hat a$ & 2.81 & 1.52&-2.98  & 1.46&  0 \\ 
$\hat c$ & 2.80 & 1.41&-2.85  & 1.43& 0.67 \\ 
$\hat b$ & 2.81 & 1.62&-3.18  & 1.56& 0.55 \\ 
\hline
\multicolumn{2}{c}{Exp. \cite{Menkovsky,Boulet}}& & & 1.42 \\
\end{tabular}
\label{tab1}
\end{table}

\begin{center}
\begin{figure}
\label{csf}
\centerline{\psfig{file=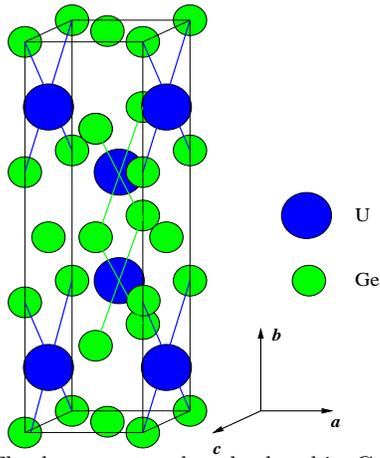,width=5cm,height=6cm}}
\caption{The base centered orthorhombic $Cmmm$ crystal structure of UGe$_2$.
The volume shown includes two primitive cells.}
\end{figure}
\end{center}

\begin{figure}
\centerline{\psfig{file=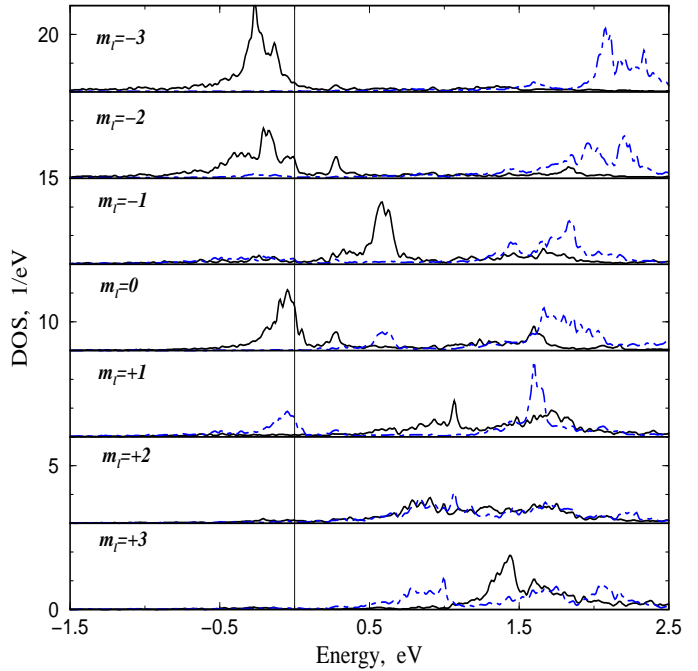,width=09cm,height=09cm}}
\caption{Spin and orbitally resolved U 5f state density of states for
easy axis $\hat a$ orientation: 
spin-$\uparrow$ is shown in black full line,
spin-$\downarrow$ is in blue dashed line.  Other orientations of the moment give 
distinctly different results.}
\end{figure}
\end{multicols}

\begin{figure}
\label{fermi}
\centerline{\psfig{file=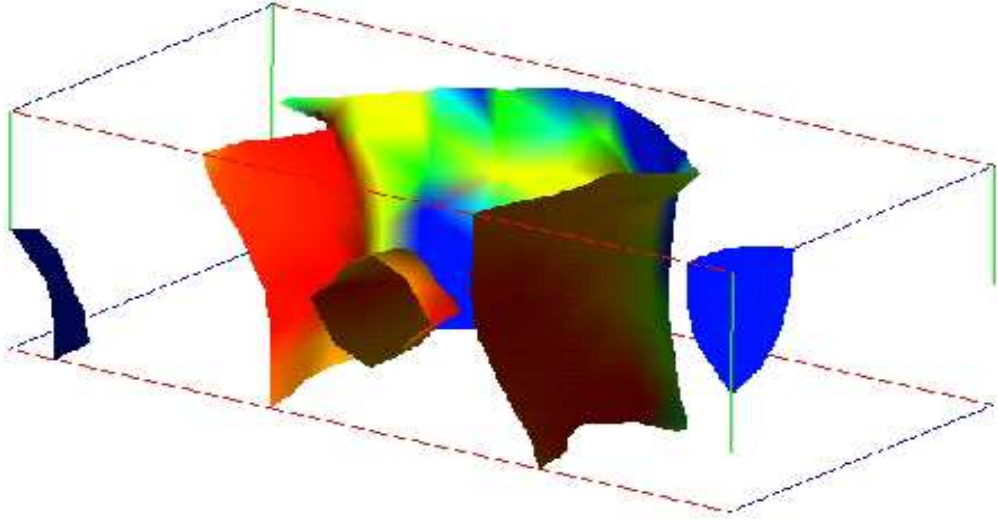,width=15cm,height=10cm}}
\caption{The Fermi Surface for UGe$_2$ that is discussed in the text.
The zone center is at the far corner of the plotted regions, and the
distances plotted correspond to $\frac{2\pi}{a}, \frac{2\pi}{b},
\frac{\pi}{c}$ with axes directed as in Fig. 1.
The large surface is primarily $\{m_s= \uparrow; m_l=0 \}$ character, is quasi-
two-dimensional and shows indication of a nesting feature around 
($0.45 \frac{2 \pi}{a}$,0,0).}
\end{figure}


\begin{references}
\bibitem{OrbMom} The orbital moment is calculated as usual from the LDA 
one-electron wavefunctions and has a different standing within spin-only
density functional theory than the spin moment.
\bibitem{CDF} 
The LDA+U uses the spin- and orbital occupation matrices to
create in addition to the spin-only dependent LDA potential, the spin- and 
orbitally dependent ``$+U$" on-site potential which produces
the orbital polarization.
\end{references}
\end{document}